\documentclass[journal]{IEEEtran}
\usepackage{cite}
\usepackage{amsmath,amssymb,amsfonts}
\usepackage{algorithmic}
\usepackage{graphicx}
\usepackage{textcomp}
\usepackage{xcolor}
\usepackage{color}
\usepackage{soul}
\usepackage{booktabs}
\usepackage{url}
\usepackage{subfigure}

\begin{document}

\title{A Survey: Towards Privacy and Security in Mobile Large Language Models}

\author{Honghui~Xu,
        Kaiyang Li,
        Wei Chen,
        Danyang Zheng,
        Zhiyuan Li,
        Zhipeng~Cai ~\IEEEmembership{IEEE Fellow}
\thanks{H. Xu is with the Department of Information Technology, Kennesaw State University, Marietta, GA, USA (email: {\it hxu10@kennesaw.edu}.)}
\thanks{K. Li and Z. Cai are with the Department of Computer Science, Georgia State University, Atlanta, GA, USA (email: {\it kli16@student.gsu.edu, zcai@gsu.edu}.)}
\thanks{W. Chen and Z. Li are with Nexa AI, Cupertino, CA, USA (email: {\it {alexchen, zack}@nexa.ai}.)}
\thanks{D. Zheng is with School of Computing and Artificial Intelligence, Southwest Jiaotong University, Chengdu, Sichuan, China (email: {\it dzheng5@swjtu.edu.cn}.)}

\thanks{Copyright (c) 20xx IEEE. Personal use of this material is permitted. However, permission to use this material for any other purposes must be obtained from the IEEE by sending a request to pubs-permissions@ieee.org.}}

\mark{~IEEE~INTERNET OF THINGS JOURNAL,~Vol.~XX, No.~XX}%

\maketitle

\begin{abstract}

Mobile Large Language Models (LLMs) are revolutionizing diverse fields such as healthcare, finance, and education with their ability to perform advanced natural language processing tasks on-the-go. However, the deployment of these models in mobile and edge environments introduces significant challenges related to privacy and security due to their resource-intensive nature and the sensitivity of the data they process. This survey provides a comprehensive overview of privacy and security issues associated with mobile LLMs, systematically categorizing existing solutions such as differential privacy, federated learning, and prompt encryption. Furthermore, we analyze vulnerabilities unique to mobile LLMs, including adversarial attacks, membership inference, and side-channel attacks, offering an in-depth comparison of their effectiveness and limitations. Despite recent advancements, mobile LLMs face unique hurdles in achieving robust security while maintaining efficiency in resource-constrained environments. To bridge this gap, we propose potential applications, discuss open challenges, and suggest future research directions, paving the way for the development of trustworthy, privacy-compliant, and scalable mobile LLM systems.

\end{abstract}


\IEEEpeerreviewmaketitle

\section{Introduction}\label{sec:introduction}

The advent of mobile Large Language Models (LLMs) represents a significant milestone in the evolution of AI, enabling advanced natural language processing capabilities to be deployed in mobile and edge environments~\cite{jin2024llms,xu2024device,ZDY-ToN}. By bringing powerful AI tools closer to end-users, mobile LLMs are revolutionizing industries such as healthcare~\cite{riad2024enhancing}, finance~\cite{zhao2024revolutionizing}, and education~\cite{kozov2024analyzing} with real-time, on-device applications. These models excel in tasks ranging from conversational AI and sentiment analysis to automated decision-making and personalized recommendations, offering unparalleled convenience and functionality in resource-constrained environments.

However, the integration of mobile LLMs into sensitive domains raises significant privacy and security challenges~\cite{potluri2024sok,yan2024protecting,zou2024adversarial,heibel2024mapping}. Mobile and edge deployments often process sensitive data directly on-device, exposing them to a unique set of risks such as adversarial attacks~\cite{greshake2023not}, membership inference~\cite{duan2024membership}, and data leakage through side-channel vulnerabilities~\cite{wang2024llms}. The resource limitations of mobile platforms further complicate the adoption of traditional security and privacy-preserving mechanisms~\cite{wiest2024anonymizing,frikha2024incognitext,singh2024whispered,lin2024promptcrypt,liu2023llms}, leaving critical gaps in protection. As these technologies become more prevalent, addressing their vulnerabilities is essential to ensure user trust and data integrity.

In this paper, we provide a comprehensive survey of privacy and security issues in mobile LLMs. While mobile LLMs unlock vast potential in enhancing real-time decision-making and user interactions, their widespread adoption necessitates a robust understanding of associated risks and mitigation strategies. By systematically analyzing these challenges, we aim to highlight both the vulnerabilities inherent in mobile LLM architectures and the potential solutions to address them. This survey is particularly relevant for AI practitioners, mobile application developers, policymakers, and end-users, emphasizing the importance of secure and privacy-compliant deployments.

\begin{figure*}[htbp]
    \centering
    \includegraphics[width=\linewidth]{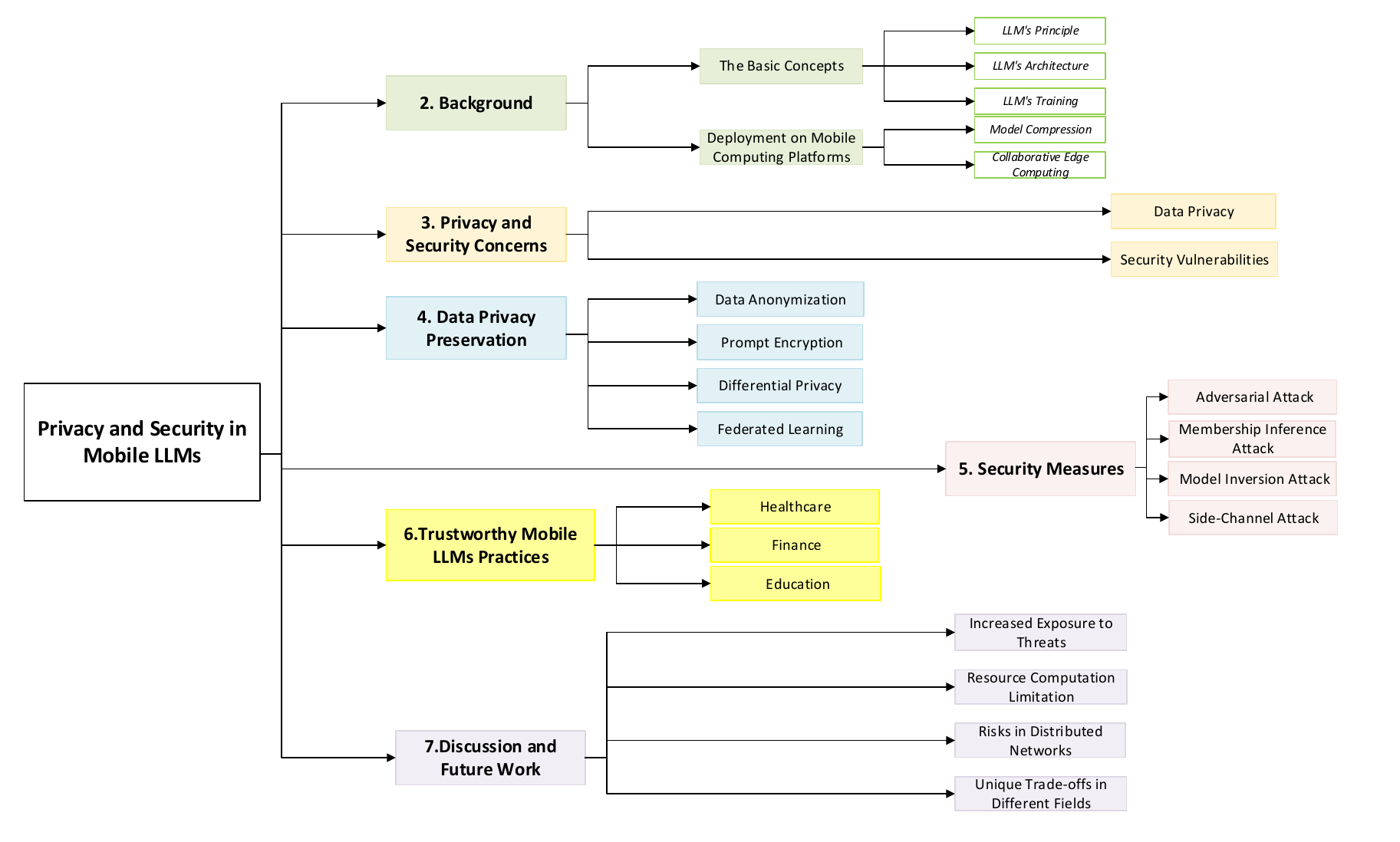}
    \caption{Overview of This Survey: Towards Privacy and Security in Mobile Large Language Models}
    \label{fig:Survey_Layout}
\end{figure*}

The remainder of this article is structured as follows: Section~\ref{sec:background} provides an overview of mobile LLMs, highlighting their capabilities and applications. In Section~\ref{sec:challenges}, we analyze the privacy and security risks associated with mobile LLMs. Strategies to mitigate these risks are discussed in Section~\ref{sec:pp_llm} and Section~\ref{sec:security_llm}. Additionally, Section~\ref{sec:case_study} presents practical examples of successful LLM implementations that uphold data security and privacy. In Section~\ref{sec:discussion_future}, we explore future directions for the secure and efficient deployment of mobile LLMs. Finally, we provide our conclusions in Section~\ref{sec:conclusion}.
The structure and scope of this survey are illustrated in Fig.~\ref{fig:Survey_Layout}.

\section{Background}\label{sec:background}

This section outlines the development, architecture, and deployment strategies of LLMs for mobile computing platforms, focusing on innovative techniques like model compression and collaborative edge computing to address performance and privacy challenges in real-time mobile applications. 

\subsection{The Basic Concept of LLMs}\label{subsec:basics}

This subsection elaborates on the foundational principles, architectural design, and dual-phase training process of LLMs.

\subsubsection{LLMs' Principle}\label{subsubsec:llm_princ}

LLMs are normally represented in the transformative advancement form in the field of NLP by offering the ability to generate, understand, analyse and manipulate the human language with unparalleled accuracy and flexibility~\cite{hadi2024large,hagos2024recent}.
These LLMs are typically built on neural network architectures such as transformers, and they are trained on vast datasets containing billions of words. This training enables them to capture complex patterns and generate human-like text. 
The core principle of LLMs lies in their ability to learn the stochastic structure of language, which empowers them to predict the next word in a sentence~\cite{raiaan2024review,yang2024harnessing}.
This capability is developed through autoregressive modeling, where the model predicts successive tokens in a sequence, allowing it to gradually learn intricate dependencies between words across diverse contexts.

\subsubsection{LLMs' Architecture}\label{subsubsec:llm_archi}

\begin{figure}[htbp]
    \centering
    \includegraphics[width=\linewidth]{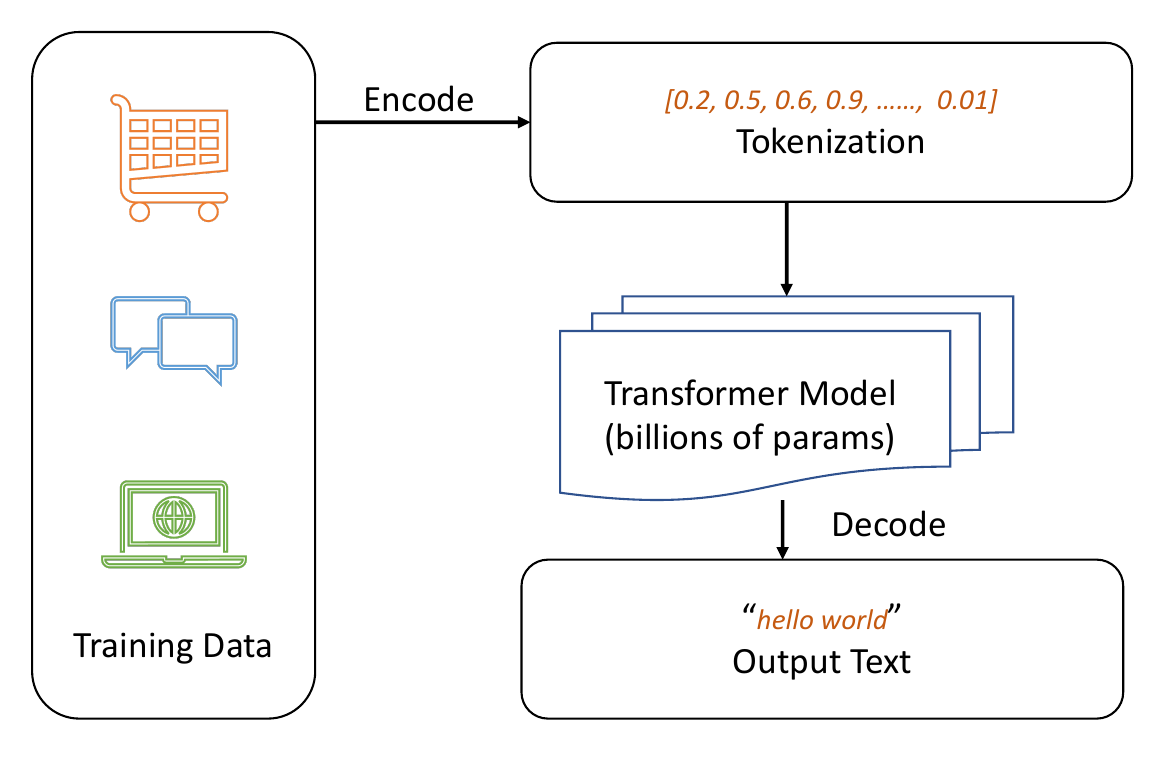}
    \caption{Data Flow in LLM's Architecture}
    \label{fig:LLM_Arch}
\end{figure}

The architecture of LLMs primarily employs transformers, introduced in 2017, which have revolutionized NLP by effectively managing long-range dependencies in text~\cite{kamath2024llms}. 
Transformers utilize self-attention mechanisms to assess the relevance of different words within a sentence for making predictions~\cite{ding2022novel}, this architecture of which comprises multiple layers of self-attention and feedforward networks shown in Fig.~\ref{fig:LLM_Arch}, enabling the model to capture both local and global contexts within text sequences~\cite{yang2021context}.
Unlike earlier models such as recurrent neural networks (RNNs) and long short-term memory (LSTM) networks, which struggled with maintaining context over longer sequences~\cite{yu2019review}, the multi-layered structure of transformers allows LLMs to model complex linguistic patterns, making them highly effective for tasks including translation, summarization, and question answering.

\subsubsection{LLMs' Training}\label{subsubsec:llm_training}

The training process of LLMs involves two critical stages: pre-training and fine-tuning~\cite{liu2024understanding}. 
In the pre-training phase, the model is exposed to vast corpora of text unsupervised, enabling it to learn general language patterns without task-specific instructions, which is crucial for developing a comprehensive understanding of language, as the model assimilates the statistical properties of text across various domains and styles~\cite{zhiyuan2024training}. 
Following pre-training, LLMs undergo fine-tuning using smaller task-specific datasets, which allows the model to adapt to particular applications, such as legal document analysis or medical text interpretation, by concentrating on the linguistic features pertinent to those fields~\cite{jeong2024fine}. 
This two-stage training approach equips LLMs to efficiently handle a broad array of NLP tasks with minimal additional training.

\subsection{LLM Deployment on Mobile Computing Platforms}\label{subsec:LL_deploy}

Recently, deploying LLMs on mobile computing platforms has become a trend, bringing advanced language capabilities to users for real-time applications like virtual assistants, personalized content generation, and context-aware recommendations~\cite{jin2024llms,xu2024device}. 
However, this deployment presents a complex set of challenges and opportunities due to the inherent constraints of mobile devices, which are primarily designed for portability, power efficiency, and connectivity, rather than the high computational power required for large, resource-intensive models like LLMs. 
Direct execution of LLMs on mobile devices without modifications can lead to slow response times, increased battery drain, and thermal management issues. 
To address these challenges, strategies have been developed to make LLMs more compatible with mobile environments.

\subsubsection{Model Compression}\label{subsubsec:model_compression}

Model compression techniques are designed to reduce the size and complexity of LLMs while preserving as much of their original performance as possible. 
Key strategies include:
(1) Pruning removes less critical parts of the model, effectively reducing the number of parameters that need processing, thereby streamlining the model~\cite{ma2023llm,cheng2024mini}.
(2) Quantization converts the precision of LLMs’ weights from floating-point to lower-bit representations, which decreases computational load and memory usage~\cite{hu2024llm,huang2024slim}.
(3) Distillation is a process, where a smaller, simpler “student” model is trained to emulate the behavior of a larger, more complex “teacher” LLM, allowing the lighter model to achieve similar performance with reduced complexity~\cite{upadhayayaya2024efficient,mcdonald2024reducing}.

\subsubsection{Collaborative Edge Computing}\label{subsubsec:cloud_infer}

\begin{figure}[htbp]
    \centering
    \includegraphics[width=\linewidth]{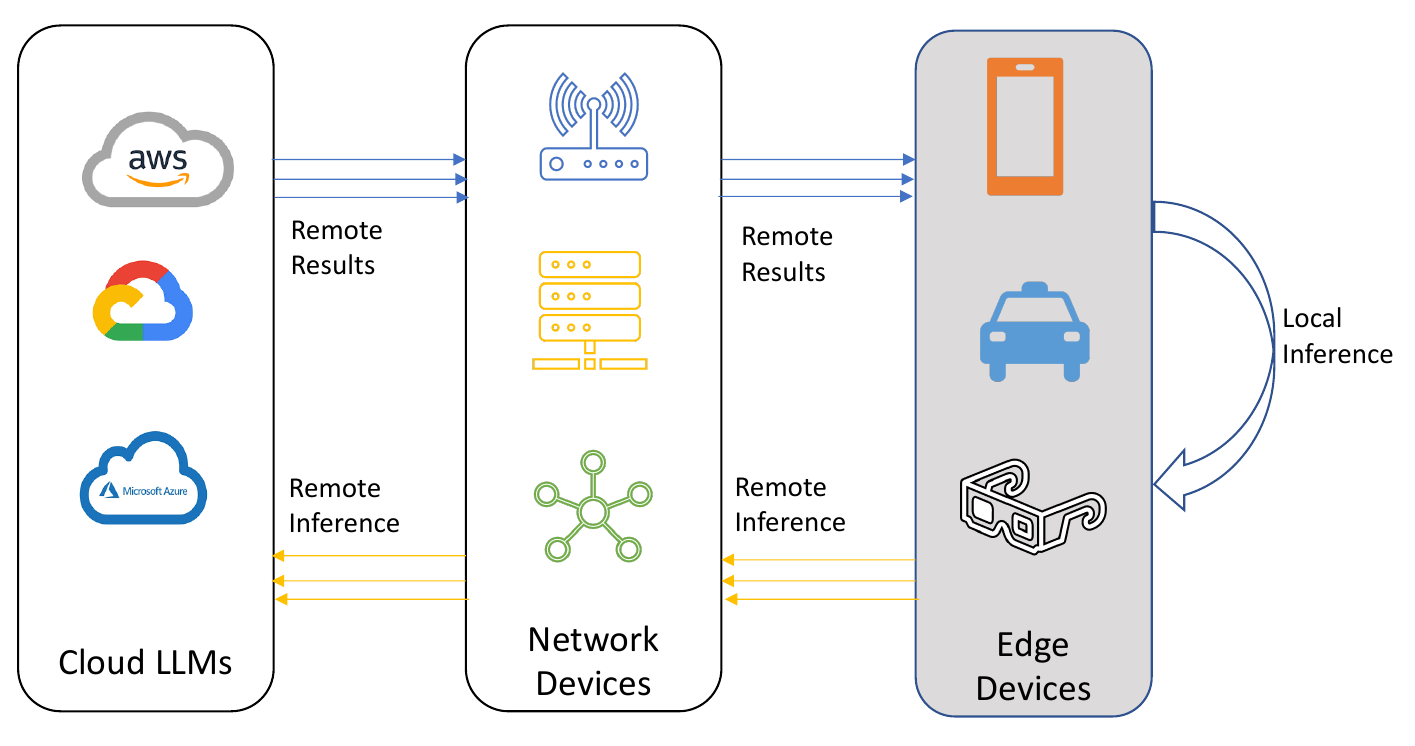}
    \caption{Collaborative Edge Computing with Cloud LLMs}
    \label{fig:collabo_edge}
\end{figure}

Another significant approach is collaborative edge computing shown in Fig.~\ref{fig:collabo_edge}, which offloads the computationally intensive tasks of LLMs to powerful edge servers~\cite{zhang2024edgeshard,ale2024empowering}.
In this setup, mobile devices transmit input data and receive processed results, significantly reducing resource strain~\cite{DBLP:journals/bigdatama/ChenHMHW24}. 
However, this strategy introduces notable privacy and security concerns since user data is transmitted to and processed on remote servers. This survey focuses primarily on examining the privacy and security challenges associated with the collaborative edge computing of LLMs.

\section{Privacy and Security Concerns for Mobile LLM}\label{sec:challenges}

As LLMs become increasingly integrated into mobile platforms, ongoing research and innovation are essential to ensure their safe and effective use. 
The deployment of LLMs on mobile computing platforms necessitates addressing critical data privacy and security vulnerabilities. 

\subsection{Data Privacy}\label{subsubsec:data_privacy}

Ensuring strong privacy regulations like the General Data Protection Regulation (GDPR) for LLMs on mobile platforms is crucial~\cite{djeki2024data}. 
Mobile devices process and often store highly sensitive personal information, including location data, contacts, messages, and browsing histories~\cite{keith2013information}. 
The risk of unauthorized access to this data is particularly concerning in mobile computing environments and could lead to significant privacy violations~\cite{zhang2018data}.
Additionally, advanced model inversion attacks add another layer of complexity by potentially allowing adversaries to reverse engineer the input data used to train the LLM, exposing sensitive information~\cite{potluri2024sok,yan2024protecting}. 
These risks necessitate securing data during processing and storage and minimizing data collection to only the essential information required for the model’s operation.

\subsection{Security Vulnerabilities}\label{subsubsec:secu_vulner}

Mobile devices are inherently more vulnerable to various cyber-attacks due to their limited computational resources and security features, compared to more robust computing environments like desktops or servers~\cite{zhang2024llms}. 
This vulnerability makes them attractive targets for attackers when using LLMs~\cite{he2024emerged}. 
For example, a significant security concern is the susceptibility of LLMs to adversarial attacks, where malicious inputs are deliberately designed to manipulate the model’s output in harmful or misleading ways~\cite{zou2024adversarial,heibel2024mapping}. 
Specifically, an adversary could craft an input that prompts the LLM to generate incorrect or biased responses. 
The risk of such adversarial attacks is especially acute in mobile environments due to the uncontrolled nature of user interactions and the varied operating conditions of mobile devices.

\section{Data Privacy Preservation for Mobile LLMs}\label{sec:pp_llm}

It is crucial to harness the benefits of LLMs without compromising personal data privacy. 
Several privacy-preserving techniques have been developed to protect sensitive user information while enabling LLMs to perform complex tasks. 
Each technique provides unique mechanisms to address various aspects of privacy concerns in mobile environments.

\subsection{Data Anonymization}\label{subsec:data_anony}

Data anonymization involves removing or obfuscating personally identifiable information (PII) from datasets to ensure privacy, a crucial process for LLMs during both training and inference stages.
By anonymizing data before training, developers prevent sensitive information from being exposed or misused. 
Recent studies have explored using LLMs to anonymize text by replacing sensitive information with non-identifiable terms, enhancing privacy without losing text readability~\cite{wiest2024anonymizing,frikha2024incognitext}. 
Other works investigated fine-tuning LLMs on datasets containing sensitive information using a dual-objective optimization framework to maintain the text’s contextual integrity while anonymizing content~\cite{singh2024whispered,li2024llm}. 
Anonymization is particularly relevant for mobile LLMs, addressing privacy concerns in mobile environments where applications access diverse user data, ensuring compliance with stringent privacy regulations and maintaining user trust.

\subsection{Prompt Encryption}\label{subsec:prompt_encry}

Prompt encryption enhances privacy in LLMs by securing prompts with cryptographic algorithms before transmission, ensuring that sensitive data remains obscured from unauthorized access~\cite{lin2024promptcrypt,liu2023llms}.
Encrypted prompts, processed in a secure environment, prevent direct interpretation by the LLM. Responses are encrypted before being sent back, maintaining security throughout the communication process. 
This prompt encryption scheme is especially vital for handling sensitive personal, financial, or medical data, as it combines robust encryption with secure processing to preserve privacy while enabling LLMs to provide accurate responses. 
However, the effectiveness of prompt encryption can be constrained by the resource limitations inherent to mobile platforms.

\subsection{Differential Privacy}\label{subsec:dp}

Differential Privacy applies controlled noise to datasets to ensure individual data points remain private, a method vital for training LLMs without compromising sensitive information~\cite{singh2024whispered}. 
Techniques like Differentially Private Stochastic Gradient Descent (DP-SGD) protect data during model fine-tuning by clipping gradients and adding noise to obscure individual contributions, maintaining efficacy in applications while ensuring robust privacy~\cite{garg2024task,DBLP:journals/bigdatama/YanZYL24}. 
Furthermore, advanced strategies utilize selective privacy application during multiple fine-tuning phases of LLMs to enhance the balance between privacy and utility~\cite{yan2024protecting}. 
For mobile LLMs, differential privacy plays a crucial role in safeguarding the vast data generated by mobile devices to ensure utility without sacrificing privacy.

\subsection{Federated Learning}\label{subsec:fl}

\begin{figure}[htbp]
    \centering
    \includegraphics[width=1\linewidth]{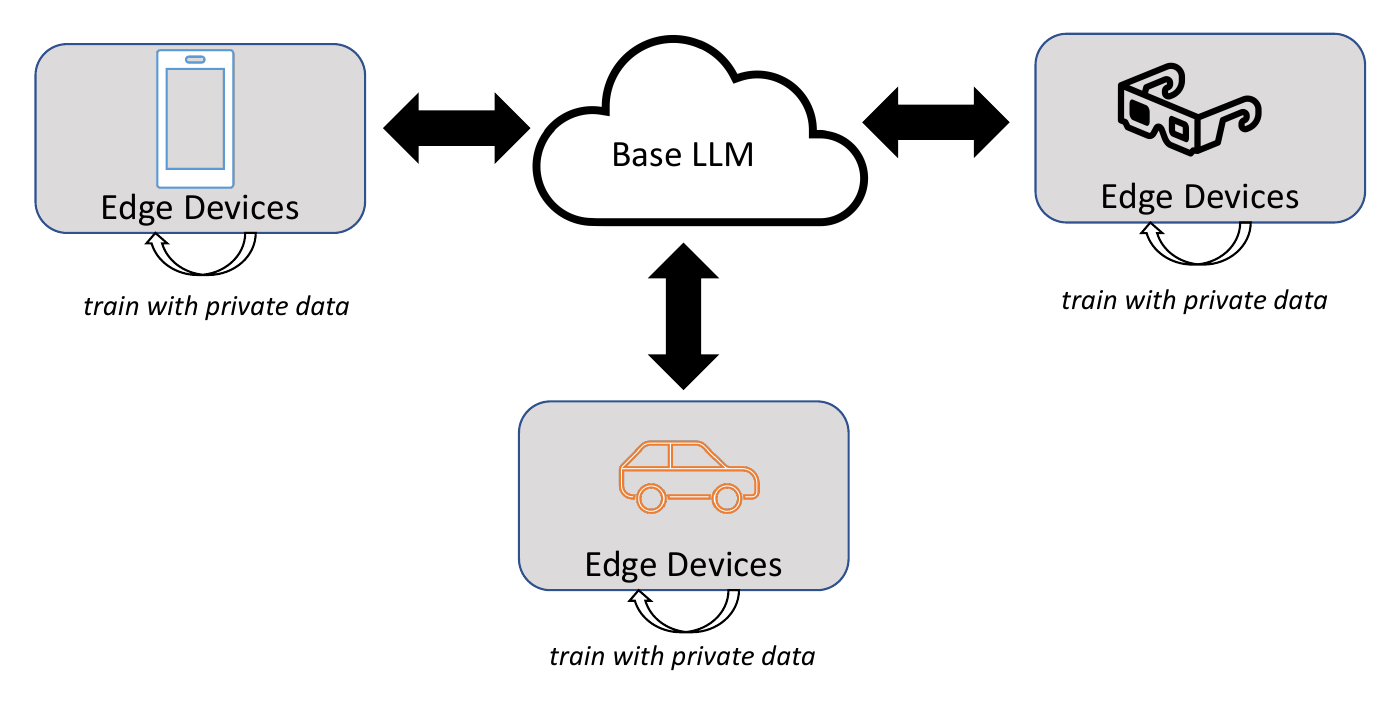}
    \caption{Federated Learning for Training Mobile LLMs}
    \label{fig:FL_LLM}
\end{figure}

Federated Learning is a decentralized method for training LLMs that enhances privacy by allowing multiple devices to process model updates locally without sharing raw data~\cite{ye2024openfedllm,kuang2024federatedscope,nikolakakis2024fedstr}, which can be found in Fig.~\ref{fig:FL_LLM}. 
Each device computes updates based on its own data, sharing only these updates, not the sensitive data, with a central server. 
This distributed computing mechanism minimizes data breach risks by keeping sensitive information on the user’s device.
Federated Learning is particularly suited for mobile platforms by leveraging their decentralized nature and computational capabilities, thus providing a privacy-preserving and efficient framework for deploying LLMs~\cite{FedBERT2022Tian,zhang-etal-2023-fedpetuning}.
Recent advancements in this domain can be categorized into four key aspects: frameworks, privacy preservation, efficiency, and handling heterogeneity.
(1) Frameworks: Frameworks like OpenFedLLM~\cite{ye2024openfedllm} and FedIT~\cite{zhang2024towards} demonstrate how FL can be leveraged for collaborative instruction tuning, improving training efficiency across diverse systems and data distributions. These frameworks establish a foundation for deploying FL-driven LLMs efficiently on decentralized platforms, such as mobile devices.
(2) Privacy Preservation: A key advantage of FL is its ability to minimize data breach risks by keeping sensitive information on user devices. Techniques like DP-DyLoRA~\cite{xu2024device} and FedPIT~\cite{zhang2024fedpit} combine differential privacy with parameter-efficient fine-tuning (PEFT) to achieve a balance between privacy preservation and computational efficiency. Furthermore, studies such as FedIT~\cite{zhang2024towards} address privacy leakage issues by mitigating data extraction attacks, ensuring the security of decentralized training environments.
(3) Efficiency: To address the computational and communication challenges of FL for LLMs, several novel methods have been proposed.  FLoRA~\cite{wang2024flora} and FedBiOT~\cite{wu2024fedbiot} utilize low-rank adaptation (LoRA)~\cite{hu2022lora} and lightweight models to reduce computational costs while maintaining scalability. Additionally, methods such as FLASC~\cite{kuo2024federated} and SplitLoRA~\cite{lin2024splitlora} focus on minimizing communication overhead through model sparsity and workload partitioning, enabling practical deployment in resource-constrained environments.
(4) Handling Heterogeneity: Heterogeneity in data and client resources poses significant challenges for FL. Solutions FeDeRA~\cite{yan2024federa} employs weight decomposition with SVD to handle non-IID data, improving performance and reducing costs. FlexLoRA~\cite{bai2024federated} allows dynamic client contributions, ensuring robust global models across diverse tasks and resource disparities. These methods highlight the importance of customizing FL frameworks to address the diverse needs of real-world deployments.

\section{Security Measures for Mobile LLMs}\label{sec:security_llm}

This section will detail the primary security measures required to protect LLMs on the mobile platforms, with a focus on defending against attack vectors, including adversarial attack, membership inference attack, model inversion attack, and side-channel attack.

\subsection{Defense on Adversarial Attack}\label{subsec:defense_adv_attack}

Adversarial attacks pose a significant threat to LLMs deployed on mobile platforms, involving the manipulation of input data to trick the model into generating incorrect or harmful outputs.
These attacks can take many forms, ranging from subtle perturbations to the input data (evasion attacks)~\cite{guerraoui2024adversarial,vitorino2024adversarial} to introducing malicious data into the training process (poisoning attacks)~\cite{zhang2024human,zhang2024instruction}. 
In mobile environments, where LLMs often rely on external data sources, such as user input or cloud-stored datasets, adversarial attacks can be particularly effective, as they exploit the dynamic nature of mobile data to bypass traditional security measures. 
For example, an attacker could modify a user’s input in a messaging app to cause the LLM to generate offensive or misleading content, potentially causing harm to both the user and the service provider~\cite{greshake2023not}. 
To defend against adversarial attacks, several techniques have been proposed, including adversarial training and model regularization. 
(1) Adversarial training involves augmenting the training dataset with adversarial examples, forcing the model to learn how to recognize and resist such inputs~\cite{fan2024towards,zou2024adversarial,zhaoenhancing}. 
While this approach has been shown to improve model robustness, it is computationally expensive and may not be feasible for deployment on resource-constrained mobile devices. 
(2) Model regularization techniques, such as L2 regularization and dropout, can also help mitigate the impact of adversarial attacks by reducing the model’s overfitting to specific input patterns, making it less susceptible to perturbations~\cite{nazari2024forget,andriushchenko2024understanding}.
However, these techniques alone may not be sufficient to fully protect LLMs from sophisticated adversarial attacks, particularly those targeting the model’s gradients.

\subsection{Defense on Membership Inference Attack}\label{subsec:defense_member_attack}

\begin{figure}[htbp]
    \centering
    \includegraphics[width=1\linewidth]{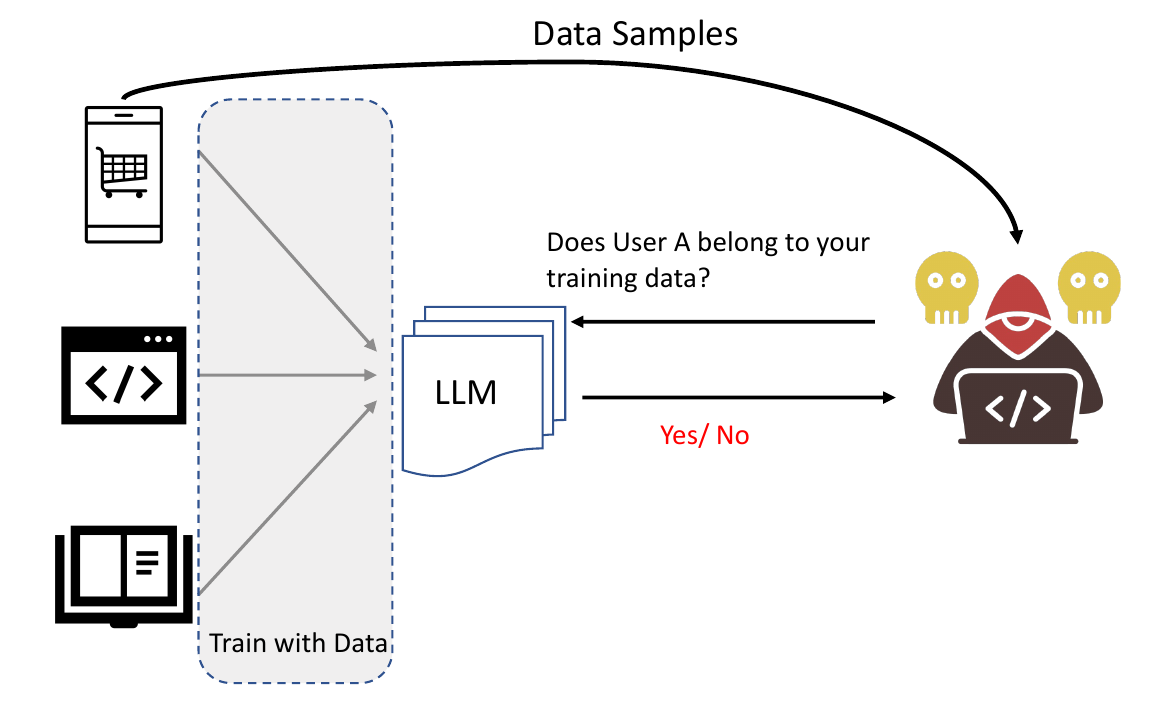}
    \caption{Membership Inference Attack on Mobile LLMs}
    \label{fig:MIA_LLM}
\end{figure}

Membership inference attacks enable an adversary to determine whether a specific data point was part of LLMs’ training datasets by analysing its behaviour during inference~\cite{duan2024membership,galli2024noisy}. 
Membership inference attacks on on-device LLMs in Fig.~\ref{fig:MIA_LLM} can have a serious privacy implications, particularly in the cases where the training data includes the sensitive information, such as medical records or financial transactions. 
To defend against the membership inference attacks on LLMs in mobile computing platforms, techniques such as differential privacy, knowledge distillation, and adaptive regularization have been proposed. 
(1) Differential privacy is effective in preventing membership inference attacks by ensuring that the presence or absence of a specific data point in the training dataset has minimal impact on the model’s outputs~\cite{lowy2024does,behnia2022ew}. 
(2) Knowledge distillation, which compresses a large LLM into a smaller, more efficient model, has been shown to reduce information leakage, as the distilled model retains only the most essential features of the original dataset~\cite{fu2023practical}. 
(3) Adaptive regularization techniques~\cite{dhingra2024protecting,halawi2024covert}, which penalize the model for overfitting to specific training data points, can further reduce the model’s vulnerability to membership inference attacks by encouraging it to generalize to unseen data. 

\subsection{Defense on Model Inversion Attack}\label{subsec:defense_inv_attack}

\begin{figure}[htbp]
    \centering
    \includegraphics[width=1\linewidth]{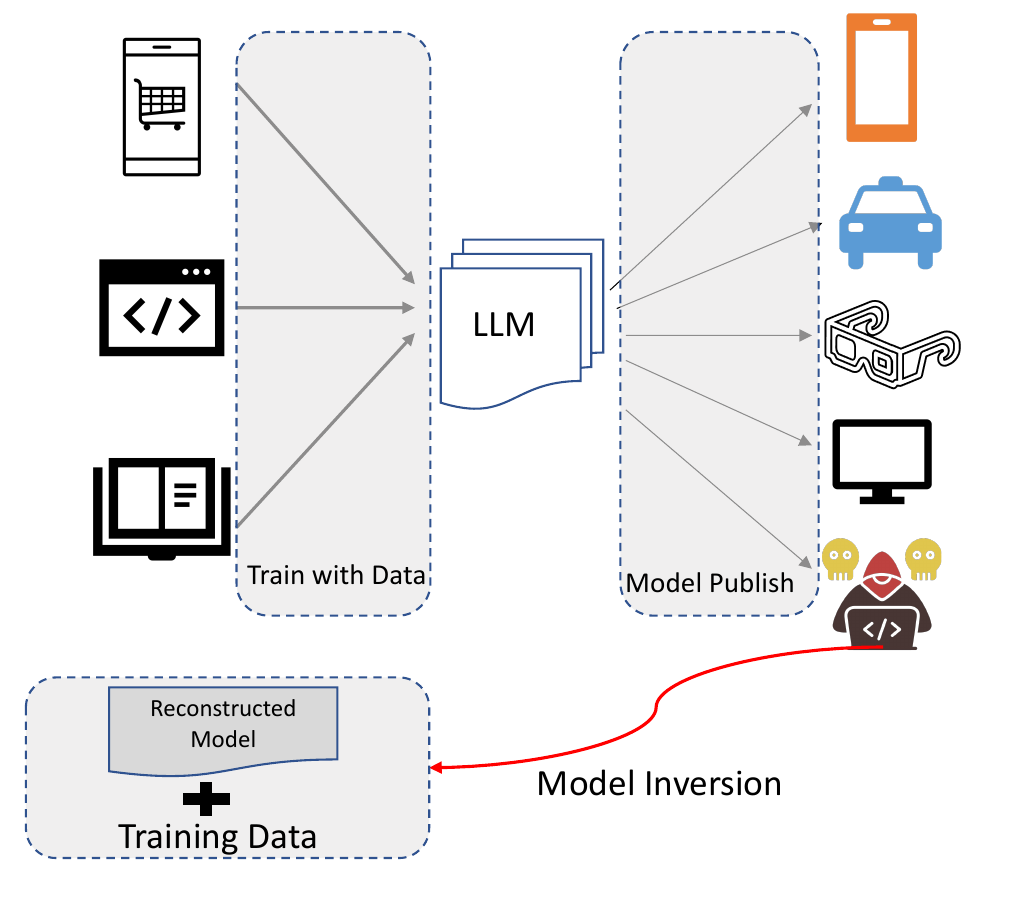}
    \caption{Membership Inversion Attack on Mobile LLMs}
    \label{fig:model_inv_llm}
\end{figure}

In addition to adversarial attacks, model inversion attacks pose a significant threat to the privacy of LLMs deployed on mobile platforms. 
As shown in Fig.~\ref{fig:model_inv_llm}, model inversion attacks allow an adversary to reconstruct the training data used to develop the model by analysing its outputs, effectively reversing the learning process~\cite{chen2024text,huang2024transferable}. 
This type of attack is particularly concerning for LLMs on mobile devices, as the training data often includes sensitive personal information, such as text messages, voice commands, or location data. 
To defend against model inversion attacks, researchers have developed several techniques, including differential privacy, secure aggregation, and noise injection. 
(1) Differential privacy introduces noise into the model’s outputs during inference, making it more difficult for attackers to infer the original training data~\cite{lowy2024does,behnia2022ew}. 
(2) Secure aggregation, which is commonly used in federated learning, ensures that model updates from individual devices are aggregated in a privacy-preserving manner, preventing any single device from reconstructing the full training dataset~\cite{guerraoui2024robust,naseri2020local}. 
(3) Noise injection~\cite{chen2023analog}, which involves adding stochastic noise to the model’s predictions, can further protect against inversion attacks by making the output predictions less deterministic, thereby reducing the attacker’s ability to infer the training data.

\subsection{Defense on Side-Channel Attack}\label{subsec:defense_side_attack}

\begin{figure}[htbp]
    \centering
    \includegraphics[width=1\linewidth]{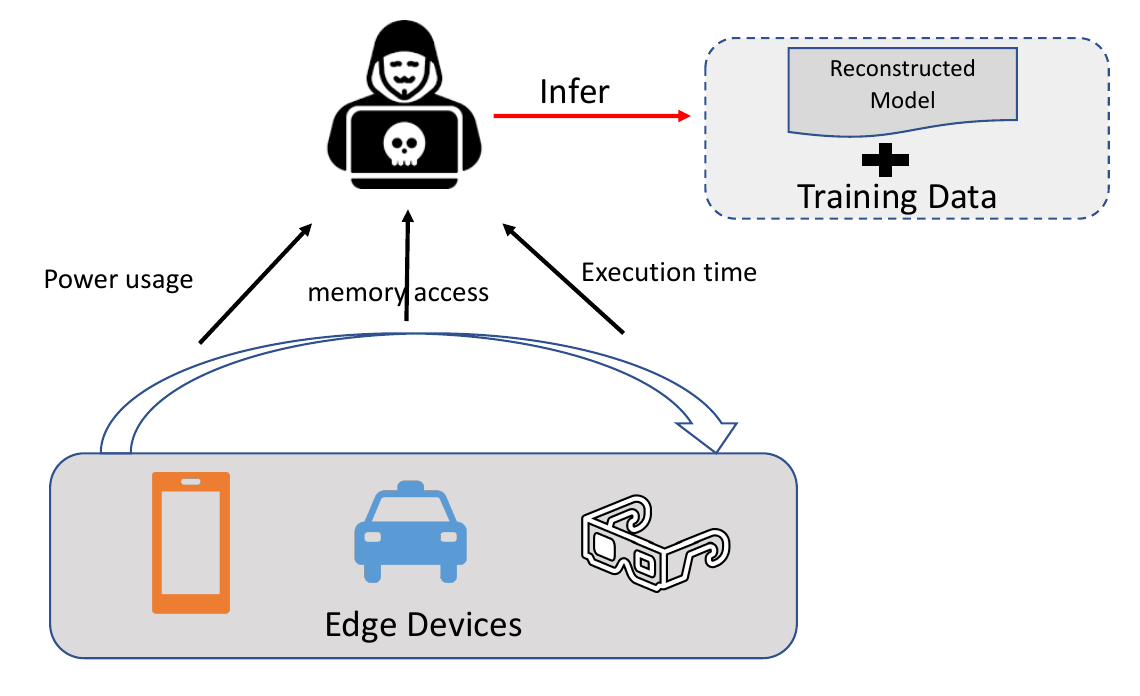}
    \caption{Side-Channel Attack on Mobile LLMs}
    \label{fig:side_attack_llm}
\end{figure}

The computational demands of on-device LLMs render them vulnerable to side-channel attacks like shown in Fig.~\ref{fig:side_attack_llm}, which will exploit physical properties such as power usage, memory access patterns, or execution times to infer sensitive information being processed by the model~\cite{wang2024llms}. 
(1) One of the most effective mitigations against the side-channel attacks is differential privacy~\cite{nazari2024llm}, a statistical framework that injects noise into the computations, reducing the correlation between inputs and outputs, thereby obscuring sensitive information from an attacker. 
Differential privacy has proven effective in mobile applications that require personalized interactions with the LLMs, such as chatbots or recommendation systems. 
(2) Additionally, federated learning with the decentralized training of LLMs across multiple mobile devices without centralizing user data will reduce the likelihood of side-channel data breach~\cite{cai2024llmaas}.
What's more, the combination of these approaches can help mitigate to many security risks posed by the mobile LLMs while preserving computational efficiency.

\section{Trustworthy Mobile LLMs Practices}\label{sec:case_study}

\begin{figure}[htbp]
    \centering
    \includegraphics[width=1\linewidth]{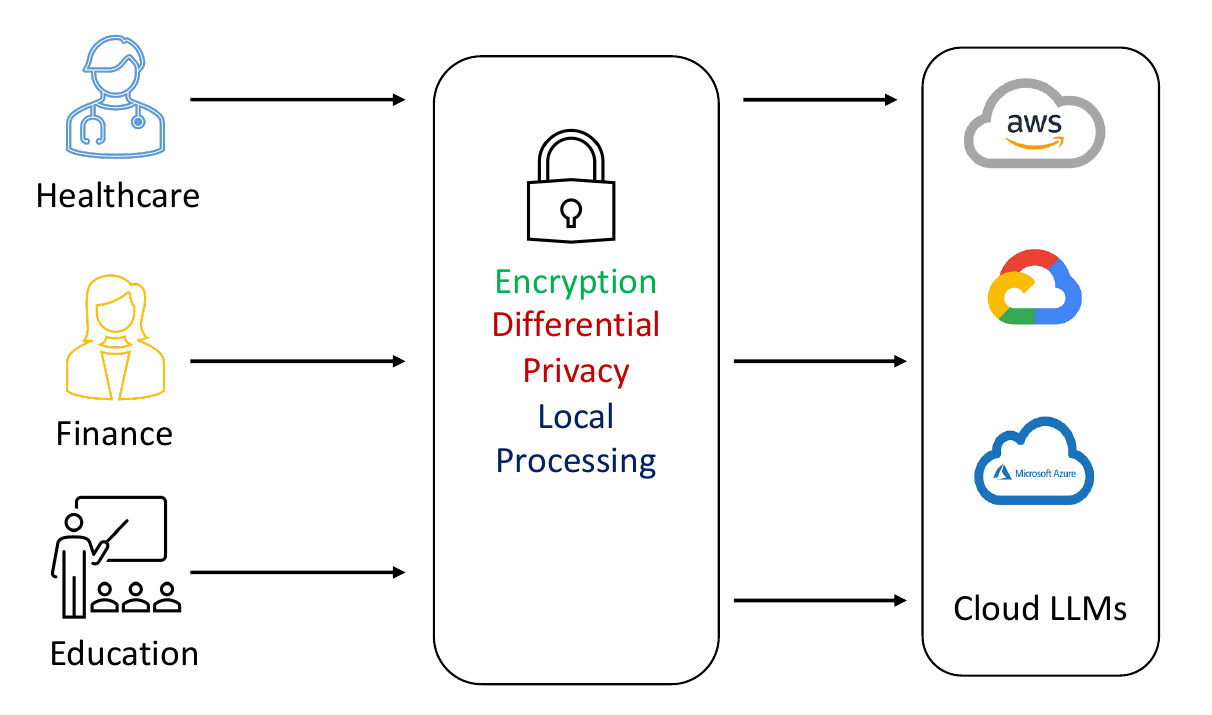}
    \caption{Trustworthy Mobile LLMs in Different Applications}
    \label{fig:trust_llm_app}
\end{figure}

In this section, we conduct a thorough examination of mainstream industries, including healthcare, finance, and education, that have successfully implemented LLMs while maintaining data security and privacy. 
These practical examples highlight the challenges of balancing performance with privacy safeguards and provide stakeholders with valuable insights into effective deployment strategies.
We will provide a more detailed explanation of trustworthy mobile LLM applications, as illustrated in Fig.~\ref{fig:trust_llm_app}, in the sections that follow.

\subsection{Trustworthy Mobile LLMs in Healthcare}\label{subsec:pp_llm_heathcare}

The integration of LLMs in healthcare has revolutionized medical data management, analysis, and patient care~\cite{de2024assessing,prabhod2023integrating}. 
Research highlights the critical role of trustworthy LLMs in mobile health applications (mHealth), which offer personalized diagnostics, treatment recommendations, and health monitoring~\cite{riad2024enhancing}.
While methods like differential privacy protect data effectively, they can impair model performance by adding noise to sensitive healthcare datasets~\cite{lin2016differential,alishahi2022add}. 
Training LLMs on heavily obfuscated datasets can lead to underperformance in crucial tasks such as diagnosis prediction or drug interaction analysis.
Although fine-tuning techniques help mitigate some performance issues, the balance between privacy and accuracy remains challenging. 
For example, in applications requiring precise patient monitoring, slight performance losses could result in serious consequences, such as delayed treatment or misdiagnosis. 
Additionally, deploying trustworthy LLMs on mobile platforms can burden system resources, as local models must manage both privacy tasks and the analysis of health data.

\subsection{Trustworthy Mobile LLMs in Finance}\label{subsec:pp_llm_finance}

The integration of LLMs into mobile applications for personal finance management and statutory auditing has notably improved decision-making, risk management, and customer service~\cite{zhao2024revolutionizing,toumeh2024assessing}. 
These applications process extensive financial data such as transaction histories, account balances, and credit scores, necessitating stringent privacy measures to prevent unauthorized access and data breaches. 
Techniques like encryption~\cite{guan2024intelligent} and differential privacy~\cite{cartwright2024evaluating} help safeguard sensitive financial information, allowing these applications to scale and serve millions of users. 
Local processing~\cite{li2024trustworthy}, where the model analyzes data directly on the user’s device without transmitting it to a central server, boosts privacy and scalability, which can reduce the risk of data breaches during transmission and is particularly valuable in mobile finance applications where remote user interaction is frequent. 
However, local processing can slow response times or increase battery usage, as privacy techniques can be computationally demanding on mobile devices. 
To address this, financial applications leverage cloud-based LLMs for intensive processing tasks while employing local differential privacy for personal data analysis on the user’s device.

\subsection{Trustworthy Mobile LLMs in Education}\label{subsec:pp_llm_edu}

The introduction of LLMs has significantly transformed educational platforms by enabling personalized learning experiences that are tailored to the individual needs of students~\cite{kozov2024analyzing,alsafari2024towards}. 
Mobile learning applications employ LLMs to provide customized educational content, including explanations, examples, and exercises, specifically designed to adapt to a student’s unique learning progress, style, and interests. 
However, the protection of students’ personal information and learning patterns is paramount, especially in environments that are susceptible to privacy violations. 
Local processing techniques~\cite{zhang2024enabling}, which ensure that only essential information is shared with the LLMs, help in safeguarding privacy, but they can also limit the effectiveness of personalization due to reduced data availability. 
Despite the application of differential privacy~\cite{gursoy2016privacy} to mitigate data exposure, balancing robust privacy protections with the efficacy of personalized learning experiences remains a significant challenge in the development of smart educational applications.

\section{Discussions and Future Work}\label{sec:discussion_future}

We will explore the complex challenges of mitigating security risks and privacy concerns for mobile LLMs. 
This discussion will shed a light on research directions that tackle these unique challenges across various sectors, enabling the safe and effective use of mobile LLMs.


\subsection{Increased Exposure to Threats}\label{subsec:increase_threat}

Mobile devices are inherently more susceptible to security breaches than centralized servers, facing threats such as malware, phishing attacks, unauthorized data access, and adversarial attacks that can manipulate outputs of LLMs. 
The physical accessibility of these devices also heightens the risk of attacks that could extract sensitive data, such as model parameters or personal user information. 
Adversarial attack, which designs malicious inputs to deceive deep learning models, are a well-known threat in the domain of LLMs. 
However, most research on improving adversarial robustness has been focused on powerful computing systems, leaving the mobile domain relatively underexplored.
Therefore, the development of mobile-friendly adversarial defenses is an essential area of future research to address the increased exposure to threats when using Mobile LLMs.

\subsection{Resource Computation Limitation}\label{subsec:resource_limit}

Mobile platforms introduce unique privacy and security challenges due to their inherent hardware limitations. 
One primary concern is the lack of a standardized privacy and security framework tailored specifically for mobile environments running LLMs. 
Unlike more robust, centralized systems, mobile devices must manage decentralized, resource-constrained operations that require efficient real-time processing. 
Current frameworks often do not fully address the unique constraints of mobile devices, such as limited computational power and battery life. 
Consequently, there is a pressing need for a lightweight framework that can guide the development and deployment of LLMs on mobile devices, ensuring robust protection against privacy breaches and security threats while accommodating the computational resource limitations inherent to these platforms.

\subsection{Risks in Distributed Networks}\label{subsec:distri_risk}

The distributed nature of mobile networks significantly exacerbates privacy concerns, as sensitive data is potentially exposed during transmission across various devices and networks, causing the risk of privacy breaches. 
Moreover, the secure data transmission between mobile devices and cloud or edge servers hosting LLMs is a critical yet underexplored area. 
Mobile platforms often rely on network communication to access the computational resources needed to run LLMs, thus creating potential attack vectors for data interception and manipulation. 
Although secure communication protocols are vital for protecting sensitive data during transmission between mobile devices and servers hosting LLMs, achieving low latency and energy efficiency remains a significant challenge for the safe deployment of LLMs on mobile platforms. 
Despite its critical importance, current research has not yet fully addressed the unique requirements for secure data transmission in mobile environments.
Future research should concentrate on developing approaches for LLMs in interconnected environments, with a focus on minimizing bandwidth consumption and energy usage while upholding robust privacy protections and security guarantees.

\subsection{Unique Trade-offs in Different Fields}\label{subsec:adapt_privacy}

Across various industrial fields, the integration of LLMs on mobile platforms introduces significant trade-offs between privacy, security, and functionality. 
For example, in healthcare, while differential privacy techniques ensure patient data confidentiality, they may degrade the performance of models, impacting crucial tasks like diagnosis and treatment recommendations. 
Financial applications leverage encryption and local data processing to enhance security, but these measures can strain mobile device resources, affecting performance and user experience. 
In educational settings, local processing preserves student privacy but may limit the effectiveness of personalized learning experiences by restricting the data available to LLMs. 
Each sector faces the challenge of designing privacy protection mechanisms and security measures that not only comply with strict regulations but also align with the unique operational demands and constraints of mobile environments
Therefore, it is also necessary to propose innovative solutions that can dynamically balance these competing priorities.

\section{Conclusion}\label{sec:conclusion}

In this paper, we provide a comprehensive survey on privacy and security challenges in mobile LLMs, highlighting their unique implications and potential solutions for real-time applications across various domains such as healthcare, finance, and education. We analyze foundational concepts like LLM architecture, training processes, and deployment strategies, particularly focusing on model compression and collaborative edge computing to address the computational constraints of mobile platforms. Additionally, we explore critical vulnerabilities, including adversarial attacks, membership inference, and side-channel attacks, and evaluate advanced privacy-preserving techniques such as differential privacy, federated learning, and prompt encryption. Our discussion sheds light on the balance between privacy, security, and functionality in mobile environments, emphasizing the need for adaptive frameworks to tackle emerging threats. Through this survey, we aim to guide future research and practical implementations of trustworthy and efficient mobile LLM systems, addressing the growing demand for secure and privacy-compliant AI applications in resource-constrained settings.

\bibliographystyle{IEEEtran}
\bibliography{LLM_Survey}

\end{document}